# Post-Vaccination COVID-19 Data Analysis: Privacy and Ethics

Sankha Das
BITS Pilani, Rajasthan, India
f20190029@pilani.bits-pilani.ac.in

Amit Dua
BITS Pilani, Rajasthan, India
amit.dua@pilani.bits-pilani.ac.in

*Abstract*—The COVID-19 pandemic has severely affected the world in terms of health, economy and peace. Fortunately, the countries are trying to overcome the situation by actively carrying out vaccinations. However, like any other massive operation involving humans such as human resource management, elections, surveys, etc., the vaccination process raises several questions about citizen privacy and misuse of personal data. In most of the countries, few attempts have been made to verify the vaccination statistics as reported by the health centers. These issues collectively require the solutions of anonymity of citizens' personal information, immutability of vaccination data and easy yet restricted access by adversarial bodies such as the government for the verification and analysis of the data. This paper introduces a blockchain-based application to simulate and monitor the vaccination process. The structure of data model used in the proposed system is based on the IEEE Standard for Data Format for Blockchain Systems 2418.2TM-2020. The proposed system enables authorized stakeholders to share and access relevant information for vaccination process chain while preserving citizen privacy and accountability of the system. It is implemented on the Ethereum blockchain and uses a Python API for the simulation and validation of each step of the vaccination process.

## I. INTRODUCTION

The COVID-19 pandemic has taken a toll over the world in several aspects over the past several months. Many lives have been lost, while many others have been severely affected from an emotional, financial and social perspective. However, there have been active efforts made in the field of medical sciences throughout this period to restore the world from the present situation back to normalcy. Many vaccines have already been developed and tested by different agencies, and several countries across continents have been actively administering these to their citizens. Most developed countries have a central system administrated by the government to define and monitor the vaccination process. The following subsections present the existing system of the vaccination process as seen in most countries and analyses it from the aspects of speed, efficiency, privacy and data integrity. These observations will thus help to lay the ground for the solution introduced in this paper.

### A. Existing Vaccination System

In order to track the vaccination of an entire country, often with population in the order of millions, a centralized and robust system is employed by the governments. Most of these systems are in the form of a portal with a graphical user interface for the citizens to register themselves into system and for the health centers to input vaccination data of the vaccinated people. These systems maintain consistency and durability of the norms and data for the process using a centralized server which is connected to the end-users by means of an application program interface (API). Different endpoints of the API give access and capture data in a specified format (most commonly JavaScript Object Notation or JSON) to and from the end-user. Almost all the endpoints have some level of access restrictions laid down by the system administrators so that certain data such as citizen information and vaccination statistics can only be read by authorized agencies such as the health centers or the government.

The first step of the process requires a citizen to register themselves on the portal. The citizen shares their personal details on the registration page such as name, age, gender, home address, etc. In addition, they are required to enter a unique number assigned to them by the government that is linked to their citizenship. Some countries call this as the Social Security number while in many others it is known as the Unique Identification Number (UUID). This number uniquely identifies the citizen and often becomes the basis of the primary key for the citizens' information in the database. Many systems automatically populate citizens' data from the government database based on the entered UUID, thus eliminating the need to enter personal details separately. Once the citizen is registered and verified, they are eligible for vaccination (subject to age and health pre-requisites) and wait for a vaccination slot.

Once a suitable slot is found at a nearby health center, the citizens go there on the allotted date and time to get the vaccine dose. Here, a verification process takes place to check if the citizen claiming to get vaccinated indeed is the one who is registered on the portal. This process is done physically by health officials often by comparing details shown in citizenship documents and identity cards with the data available on the vaccination portal. Additional measures such as biometric verification may take place but are kept to a minimum considering the pandemic situation where contact minimization is a must. Once the verification is complete and certain health checks have been done, the health officials proceed to administer the vaccine dose to the citizen. This is followed by marking the citizen as vaccinated on the portal.

The research work carried out in this report was supported by IEEE Standards Education Grant.

Information such as vaccine name, dose number, date and time of vaccination, health official's details and health status of the citizen are entered into the portal, which successfully marks the citizen as vaccinated. The access to this step is only available with the health officials so that the data entered is credible and verified. The process mostly ends with the citizen getting a copy of the certificate of vaccination which can be in the form of a hard copy or a digitally signed certificate issued by the government.

A significant portion of the vaccination process is completely physical and has little scope of digital data exchanges taking place. Physical verification of citizen identity and administering the vaccine dose is completely left to the health officials working at the health centers designated for vaccination.

### B. Analysis of the Existing System

The existing vaccination system, while being robust and sufficiently efficient to cater the management of the vaccination process in most countries, has some inherent shortcomings. This subsection draws from the different steps in the above mentioned process and identifies the drawbacks in each. At the same time, we form a basic idea of what a possible solution to the problem could be, so as to form the foundation for the solution introduced in this paper.

The process of registration of the citizens on the portal may involve making a copy of the citizens' data in the portal. This data includes personal information such as UUID, name, phone number, address, date of birth among many others. Although most of these details have little confidential value, compromise of some details such as date of birth, phone number and even home address may result in breach of privacy of the concerned individual [1]. At the same time, one may observe that a large amount of the personal data is not required for the purpose of vaccination or its analysis either. Information such as name, home address, phone number, etc. are hardly useful in the process of analysis of vaccination data. Some unique information such as phone number or UUID may be required for the purpose of verification at the vaccination centers. However, the storage and exchange of such information needs to be improved so as not to compromise the privacy of the citizen. Creation of pseudo-identities which have a one-to-one mapping with the pool of citizens could be a possible solution. These pseudo-identities would be indexed by pseudo-UUIDs which are generated through a one-way function from the actual UUID along with some randomizing factor. Hence, a non-traceable entity is created solely for the purpose of vaccination and further analysis.

The maximum scope for compromise of sensitive data lies at the actual vaccination sites. Most of the data exchanges take place during identity verification and vaccination confirmation. During identity verification, the individual is generally asked for their UUID which is queried in the portal and the returned results are compared with the individual's claims and identity cards. A large amount of personal data is exposed to a third party (the health official in this case) in this step as information such as UUID, phone number and date of birth are checked and used for verification. Information such as phone number and date of birth are often used for verification of bank account transactions and password recovery operations. This data might be maliciously stored in the background which is a major breach of privacy for the individual about to be vaccinated. In case the data is stored, it may be further compromised by sharing to other individuals, who might as well use the data in a wrong manner.

Even in this step, one can identify the need to eliminate redundant yet sensitive information such as date of birth from the verification system. As far as analysis is concerned, age of the individual is a more important metric than the date of birth, which may be calculated and stored which makes no compromises to the citizen's privacy. Verification of identity is required to be done in a more ingenious fashion which requires minimum data to shown to the health officials and most of the verification process is carried out by the citizen themselves. Therefore, existing solutions such as one-time passwords (OTP) [2] and pseudo-identities as mentioned above need to be combined to form a verification system that is more private to the concerned individual.

With the existing situation described and inherent problems in the same identified, we have formed a sufficient amount of idea as to how a more robust and privacy-preserving solution to the vaccination process can be developed. Effectively, there is a need of a system which forms a layer of abstraction between the actual citizen data and the data that is finally stored and used for further analysis. Identity verification systems need to come up with algorithms that expose just the minimum information necessary to verify the individual's identity, while most of this process being initiated and mediated by the individuals themselves. This paper addresses the design of such a system, with an added layer of a blockchain ledger to assist in the verification transactions that take throughout the registration and vaccination process. We present a decentralized system that captures the essence of vaccination information of the citizens, taking privacy and accountability of all participants in the system into consideration. The proposed system provides immutability and transparency of all transactions in the system as well as distributed trust assumption that brings consensus based transaction processing.

### C. Security and Privacy

In general terms, security is defined as a state of being in a secure and threat-free condition. The notion of security can be extended to multiple subjects such as security of life, occupation, property, monetary assets, etc. In the scope of cybersecurity, it primarily deals with security of data. Different organizations across domains have their own well-defined models and norms of security to best suit their interests.

One of the most popular notions of security is the CIA triad which stands for confidentiality, integrity and availability of data [3]. Confidentiality implies that data can be accessed and modified only by authorized users and digital processes. Integrity of data means that data is not tampered with and is

The research work carried out in this report was supported by IEEE Standards Education Grant.

not accidentally or deliberately corrupted. Lastly, availability ensures that data is accessible to the authorized users as and when required in the required form.

One can imagine security to be a intellectual game between an attacker and the user or adversarial. If the user takes appropriate measures to keep their data secure, the attacker devises new methods to gain unauthorized access to the data and often maliciously modify it as well. In response to the attack, the user tries to up their game as well by increasing the security standards of their system and the cycle continues. Security can be implemented at both hardware and software levels. While hardware security mostly deals with physical protection of devices, secure chip design and preventing side-channel attacks [4], software security is a more versatile and widely-researched topic. At a software level, people resort to preventing access to data by implementing authorized access via passwords and access tokens. As an alternative, data is also kept confidential by encrypting it such that only users who have access to the encryption key can have access to the underlying data. In the proposed system, we also use popular symmetric-key cryptography encryption algorithms such as AES.

Privacy is closely related to security and is a concept that deals with keeping the personal information pertaining to an individual or organization confidential to them. Privacy of personal data is a huge concern in the modern times with the advent of social networking sites and public information ledgers where free and unrestricted exchange of an individual or organization's data is continuously taking place [5]. This data might often be sensitive in nature such as phone numbers, date of birth and bank account details which may lead to severe loss of property and even loss of identity in some cases. Some countries have legislated privacy acts where heavy penalties and lawsuits are imposed on a person or organization who might try to compromise someone else's sensitive and private information. One such example is the Health Insurance Portability and Accountability Act of 1996 (HIPAA) enforced in the USA to protect sensitive data of a patient from being disclosed without their consent or knowledge [6]. It thus becomes a matter of ethics and morals for the individuals and firms that need access to the personal data of a person, to use it with discretion. These values are highlighted and implemented in the solution proposed in this system.

*D. Cryptographic Primitives*

Data confidentiality in the proposed system is managed by the standard symmetric key algorithm. Symmetric key cryptography involves encrypting a message with a key and the same key is also used for decrypting the plaintext message from the cipher text, thus justifying its being called symmetric [7]. Encrypting a plaintext message generally involves scrambling the message in such a manner that the cipher text doesn't make any sense to the reader. In the case of symmetric key encryption, this is achieved by using either block ciphers or stream ciphers which add confusion and diffusion to the message based on the input key [8]. This scrambling of data is one-way in nature and makes it practically infeasible to decipher the cipher text without access to the key. In the reverse process of decryption, the key is used to restore the original arrangement and phase of the data bits by using the inverse of the functions used in the encryption process. A message encrypted with a particular key cannot be decrypted from the cipher text by using any other key. This key acts as a shared secret between the trusted communicating parties which use it to encrypt and decrypt messages exchanged with each other. Therefore, only they can decrypt the cipher text as only they have access to the shared key.

Symmetric key encryption algorithms, as compared to their asymmetric key counterparts, are known to be more efficient and fast. Keys for symmetric encryption are typically of the length of 80 to 128 bits which is much smaller than asymmetric keys which are generally 1024 to 2048 bits long. Hence, symmetric keys have a storage advantage over asymmetric keys. The encryption and decryption processes are also faster in the case of symmetric key algorithms. Widely used symmetric key algorithms are the Advanced Encryption Standard (AES) [9] and Rivest Cipher 4 (RC4) [10] which are examples of block cipher and stream cipher respectively. AES is widely used across the world in defence systems, banking systems, and many other security applications.

In contrast to symmetric key cryptography, public-key (a.k.a. asymmetric key) algorithms use two different keys [11]. The keys are known as a public-private key pair and are related through mathematical functions. The private key should be kept secret to the key owner while the public key is made publicly available. A message encrypted with a user's public key can only be decrypted with the corresponding private key. On the other hand, a signature on a message is generated using the user's private key, which will get verified with the corresponding public key. Widely used public key algorithms are RSA, DSA and ECDSA [12].

A class of functions called Hash functions play a crucial role in the area of security and cryptography. Hash functions are used to generate fixed-length message digests of data which are widely used for data integrity verification, password storage and version control [13]. A cryptographically secure hash function should satisfy the properties of collision-resistance and one-way mapping. Collision resistance means that it should be significantly difficult in terms of time and resources to find two distinct inputs that produce the same output when passed as arguments to a hash function. Mathematically, it is infeasible to find $x$ and $y$ such that $x = y$ and $H(x) = H(y)$, where $H$ is the hash function. Being one-way in nature means that given an output $y$ of a hash function $H$, it is infeasible to find $x$ such that $H(x) = y$. Due to these properties, hash functions play a pivotal role in security applications and are one of the most important pillars of blockchain technology as described in the next subsection.

*E. Blockchain Technology*

Blockchain has been one of the most widely emerging technologies over the past one decade. A blockchain is a distributed and immutable ledger of transactions [14]. It was introduced as the underlying technology for the Bitcoin cryptocurrency in 2008 in the Bitcoin whitepaper [15]. A blockchain, as the name suggests, can be imagined as a chain of blocks containing data. It is essentially implemented as a linked list with each block of data containing a pointer to its previous block in the chain. The pointers in the case of blockchain are special in the sense that they are hash pointers. Hash pointers point to the previous block and also store the hash or digest of the data contained in the previous block. The data contained in a block in most blockchains generally contains information such as timestamp of its creation, block size, block address, and a stack of transactions organized as a special data structure called a Merkle Tree [16]. These transactions are proof of a certain operation having taken place at some time in the future and cannot be mutated. The immutability of the data in a block is due to the fact that changing the data in one block changes the hash value of the block. This creates a conflict with the hash value stored in the next block's hash pointer, thus creating a ripple effect through the blockchain.

The blockchain ledger is distributed across a peer-to-peer network of computers, also called as nodes [17]. Every node preserves a local copy of the blockchain and always maintains the latest state of this copy. Every new transaction that is added to a block in the blockchain is first verified by each node in the network. In most blockchains, a cost is incurred with each transaction which is paid by the transaction initiator. This cost is used to pay the block miner, which is computer on the network which invests time and resources to create a new block or add a transaction to the block. The first block or genesis block is created by the blockchain owner and has a block height of 0. The block height of a block is defined as its distance from the genesis block in terms of intermediate blocks. In order to mine a block, a node has to solve a cryptographic puzzle which is time and resource intensive to solve. The node which solves the puzzle first gets to create the block and adds it to the end of the existing chain. In case two nodes solve the puzzle at the same time, then two blocks are created. One of the block gets added to the primary chain and the other one gets added as an orphan block that exists outside the main chain. All the newly added blocks point to the previously existing terminal block as their previous block and store its hash value in their hash pointers.

After seeing the enormous success of Bitcoin in the financial industry, blockchain has found enormous acceptance in applications such as healthcare, supply-chain, asset management and so on. In addition to Bitcoin, there are other blockchain platforms like Ethereum, Hyperledger Fabric, etc. which have adopted in industry working based on various consensus protocols and smart contracts. In the proposed system, Ethereum platform is considered due to its enhanced capability with respect to consensus protocols and smart contracts [18].

The remainder of this paper is organized as follows. Section II lays out the basic design of the proposed system along with the interacting models and entities. Section III talks about the various algorithms that are used in the system and interact with the system entities. Section IV describes how the project uses the IEEE Standard for Data Format for Blockchain Systems followed by the implementation sketches of the developed application. Section V concludes the work highlighting some prominent future scope of the research work which may carried out through this work.

## II. SYSTEM MODEL DESIGN

In order to understand the functioning of the proposed system, it is necessary to first take a deeper understanding of the design of the models and database entities that will be involved in the system. The realization of the vaccination process using the proposed system is achieved through the interaction of these entities with the different algorithms and protocols designed to achieve the objectives of the real-world scenario. At the same time, it is necessary to build the objects in a manner that is compatible with the targets of user data privacy and ethical verification of vaccination statistics. The following subsections describe each of the required models that will be interacting in the simulation of the proposed solution.

### A. Government Agency

The Government Agency is the master entity in the proposed solution. Taking the cue from the real-world scenario, most of the process underlying the vaccinations will be decided by the government. It is the government that will be licensing health centers to act as authorized vaccination centers. Since a single government agency as a centralized power would overload it with the responsibility of overseeing thousands of vaccination centers in the country, it is an efficient move to divide the power of the single government into multiple acting agencies. This in turn adds the benefits of decentralization which will be of paramount importance to our solution as it needs to align with the core principles of the blockchain technology. The formation of agencies can be done on a regional basis, with each agency taking the charge of overseeing the process in a particular PIN code area or district.

A government agency will have certain top-level permissions to authorize and access the other entities in the system. For example, the system requires each participating entity to own a certain static key, which will be used in the forthcoming steps for the purpose of identity verification and authorization. For purposes of stability of this process, it is required that only limited keys are generated by a signing authority so that spurious or duplicate keys are not floated in the system. In our use case, it is best to wrest this authority with the government agencies. During the process of data analysis, it is the government agencies that will be able to access the relevant data and verify it with the transactions registered in the blockchain.

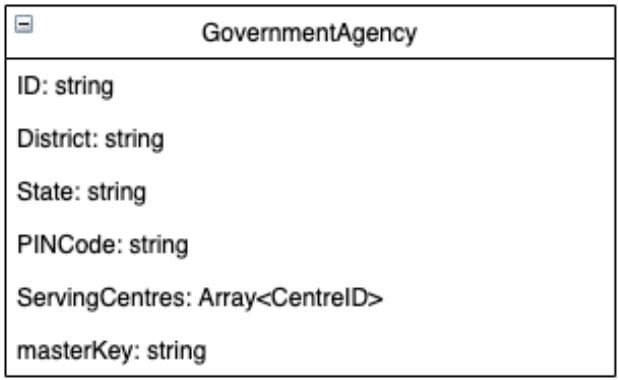

Fig. 1. Government Agency Entity Model

## B. Vaccination center

Vaccination centers are the actual sites of vaccination of the citizens. As explained above, vaccination centers carry out citizen identity verification prior to administering the dose. Therefore, it is required that they are given certain permissions to access citizen data but in a more restricted manner. Post-vaccination formalities such as confirming the vaccination of the citizen is also carried out by the health officials at these centers through the same portal. In such a case, it is required that they act as an authorized signing entity to confirm the vaccinations.

In order that new vaccination centers are not created on the fly, it is required that their authority is handed out only through the government agency acting in the area. The signing credentials of the center such as its identity number and static key will be derived from the government agencies' master key. All certificates of the vaccinations carried out at a center will be digitally signed jointly by the center and the corresponding citizen to maintain a transparent system of vaccination.

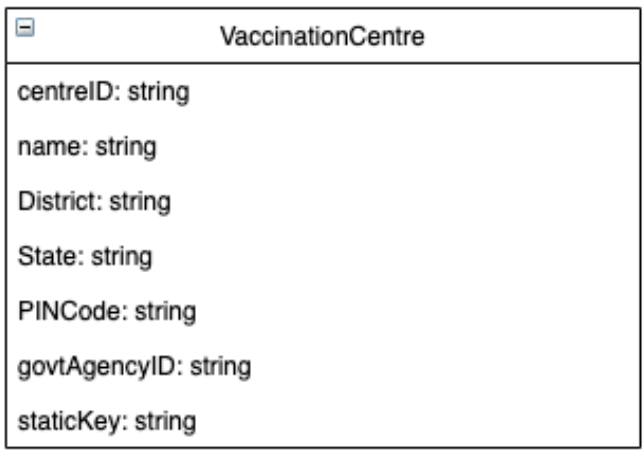

Fig. 2. Vaccination center Entity Model

## C. Citizen or User

The citizens are the primary subjects of the system whose data needs to be handled in a delicate and secure manner while registering them on the system. At the same time, personal data needs to be duly verified in order to register and vaccinate them. Although the citizens may choose to get themselves vaccinated at any center of their choice, they will be registered under and associated with a government agency closest to their current place of registration. This information will also be used to create their pseudo-UUIDs and static keys which will be in part derived from the master keys of the associated government agencies. As will be seen in the forthcoming sections, the static key of the citizen will be used to generate verification pages for instant identity verification at the time of vaccination. Once a citizen is vaccinated, the confirmation of the same will be jointly signed by the citizen and the vaccination center.

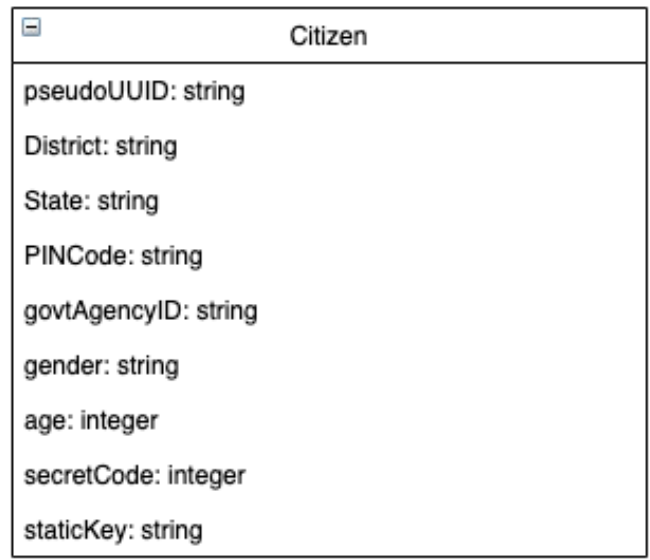

Fig. 3. Citizen Entity Model

The relevant data for a citizen first needs to be identified which will be indispensable in the stages of verification and vaccination analysis. This may require deriving and using only required data from the stored sensitive information while registering the citizen on the portal. For example, a citizen's date of birth may be a sensitive information that should not be revealed to third parties, but the age of the citizen is an important metric for the purpose of post-vaccination analysis. Therefore, at the time of registration, the age can be calculated from the date of birth and stored, and the date of birth will no longer be stored or exchanged in the portal. Apart from the targets of privacy and ethics, certain information needs to be stored for the purpose of system performance optimization. For instance, the government agency might want to verify vaccination statistics per district, and for this needs to query all stored data about the citizens from that district. Using a naïve linear search through the database will render the query slow and inefficient. On the other hand, one may create a search index on details such as district, state, age, etc. to optimize such queries. Therefore, the solution also proposes to store details such as district, state, PIN

Code, age, etc. which are not uniquely re-traceable to the citizens and are also useful for other operations.

Fig. 4. Screenshot of the application with a newly registered citizen's pseudo- identity data

*D. Vaccination*

For the purposes of further analysis and verification of the vaccination process, there is a need of storing important details associated with each vaccine dose administered to the citizens. Therefore, creating a separate entity in the system makes the access to the required data easy and also enhances system performance in terms of speed and efficiency, although with an added cost of storage space. To make the best compromise between storage and performance, one may choose to include only the most important details about the vaccinations to be stored in the database. Aligning with the pillars of privacy and ethics, these details should not reveal any sensitive data about the vaccinated citizen.

Fig. 5. Vaccination Entity Model

We have already presented a high level overview of the system entities required for the simulation of the proposed solution. These models will be used to create transaction models as described in section IV to store the process history in the blockchain. At the same time, the described models will be used as a reference in the next section which discusses the algorithms employed in the procedures involved in the system.

Fig. 6. Screenshot of the application showing details of a citizen's vaccine dose

III. PROPOSED SYSTEM

The proposed solution aims to emulate the real-world vaccination process with added features of privacy and history verification using a blockchain. However, the underlying basic processes such as citizen registration, identity verification and vaccination confirmation still remain the same. The entire process has been divided into three main stages: citizen and center registration, identity verification and vaccination confirmation. Each of these stages are further divided into atomic steps which have their dedicated set of algorithms. The system assumes that symmetric and private keys of users must be kept secret with the user, whereas, public key along with associated public parameters are made public. Also, smart contracts and consensus protocols of the Ethereum blockchain support intended security features.

*A. Center Registration*

The first step in setting up a new vaccination center is to register it on the system portal along with authorizing the center for vaccinations. An existing center creates a profile on the portal by giving details such as its name, location, center personnel, etc. The center's location details are used to find the nearest government agency, and the center is thus associated with this government agency.

Once the center has been linked to the government agency, additional details specific to the system are generated for the center. Every new center is assigned an ID number which is derived from its basic details such as name, address and an additional parameter based on the associated government agency's master key. Once the ID number has been generated, a static key for the center is generated which is again derived from the government agency's master key. The master key is used as an additional authorizing measure so that new centers cannot be registered randomly without any restriction. At each generation step randomizing factors based on current timestamp are also added so as to make the credentials as unique as possible. In each procedure, cryptographic hash functions such as the SHA-256 function are used so as to prevent the extraction of sensitive data like the government agency's master key through reverse-engineering.

The below written pseudo-codes for the associated methods summarize the steps involved in registering a new center on the portal and the process of authorizing the same.

**Algorithm 1** Center(PINCode: string, address: string)

*govtAgency ← GovtAPI.getGovtAgency(PINCode)*
*this.centerID ← generateID(masterKey, address)*
*this.PINCode ← PINCode*
*this.address ← address*
*this.staticKey ← generateStaticKey(*
    *govtAgency.masterKey, this.centerID)*

**Algorithm 2** completeCenterRegistration(response: map):

*this.centerName ← response.name*
*this.district ← response.district*
*this.state ← response.state*

### B. Citizen Registration

The citizens' registration involves a few more intermediary steps as compared to the center registration. A citizen is required to register on the portal by entering their UUID and phone number. If the UUID exists on the government's records, then an OTP is sent to the phone number entered to verify the authenticity of the citizen. The citizen is meanwhile directed to another page on the portal to enter the OTP. If the entered OTP passes the challenge, then the next steps for citizen profile creation begin.

First, a pseudo-UUID is generated from the original UUID and a timestamp based randomizing factor. Here we use a double SHA-256 hash of the concatenated UUID and randomizing factor as the pseudo-UUID. The randomizing factor ensures that a brute force search on all existing UUIDs cannot be applied in order to derive the UUID from a specific pseudo-UUID. The generated pseudo-UUID becomes the new unique identifier of the citizen for all purposes in the vaccination system, including vaccine administering and analysis. Other basic details such as name, age, gender, district, state, etc. are populated from the government's UUID database and the citizen's response into the citizen's profile. At this stage, original UUID is dropped from the system and all further operations are made on the pseudo-UUID only.

Once the basic details about the citizen have been added to their profile, the citizen is linked with a government agency closest to their current address. After this, a static key is generated for the citizen in a process similar to that for the vaccination centers using the associated government agency's master key. In addition, a secret code is generated and assigned to each citizen. The secret code is derived from the citizen's pseudo-UUID and their PIN code. This additional constraint is imposed so that no two citizens living in the same PIN Code area (and hence associated with the same government agency)

**Algorithm 3** generateStaticKey(masterKey: string, centerID: string):

*r ← random(seed = utc.now().seconds())*
*staticKey = SHA256(XOR(centerID, masterKey, r))*
**return** *staticKey*

can have the same secret code. This secret code will be used in the steps of pre-vaccination identity verification as described in the next subsection.

**Algorithm 4** Citizen(uuid: string):

*r ← random(seed = utc.now().seconds())*
*this.pseudoUUID ← generatePseudoUUID(uuid, r)*
*this.age ← calculateAge(GovtAPI.getDOB(uuid))*

**Algorithm 5** completeRegistration(response: map):

*this.PINCode ← response.PINCode*
*this.district ← GovtAPI.getDistrict(this.PINCode)*
*this.state ← GovtAPI.getState(this.PINCode)*
*govtAgency ← GovtAPI.getGovtAgency(this.PINCode)*
*this.staticKey ← generateStaticKey(this.pseudoUUID,*
                *govtAgency.masterKey)*
*this.secretCode ← generateSecretCode(*
         *this.pseudoUUID,*
         *this.PINCode)*

**Algorithm 6** generatePseudoUUID(uuid: string, r: integer):

**return** *SHA256(SHA256(uuid(.)r)*

**Algorithm 7** generateStaticKey(pseudoUUID: string, masterKey: string):

*r ← random(seed = utc.now().seconds())*
*staticKey ← SHA256(XOR(pseudoUUID, masterKey, r))*
**return** *staticKey*

### C. Pre-Vaccination Identity Verification

Once a citizen has been registered on the portal, they are eligible to get vaccinated, provided they do not have any health restrictions otherwise. A person goes to the nearest registered vaccination center to get a dose of the available vaccine. As can be seen in the case of most COVID-19 vaccines, a person needs to be administered 2 doses of a vaccine in order to be fully vaccinated. Thus, during each vaccination, the person is also checked for the number of previous doses administered to them apart from verifying their identity.

The identity verification process takes place at the vaccination center prior to the vaccination. The health officials present at the site follow a set of standard procedures to check if the person's identity claims match with the government records. Generally the citizen presents documents such as a government-issued identity card and their details are also queried on the vaccination portal to match the two sets of details. However, this process has a risk of exposing sensitive information about the citizen such as date of birth and phone number to the health officials or other people present at the site. The proposed solution seeks to tackle this problem by

**Algorithm 8** generateSecretCode(pseudoUUID: string, PIN-Code: string):

    **return** $ceil(log_2(SHA256(pseudoUUID(.)PINCode))) * 5$

reducing the amount of data that is needed by the health officials to completely verify the identity of the citizen, while at the same time increasing the involvement of the citizen in the verification process.

**Algorithm 9** vaccinate():

    $response \leftarrow HTTPResponse()$   //center response
    $center \leftarrow Center.queryset(response.centerID)$
    **try** $vaccination = createVerificationPage($
      $response.secretCode, response.PINCode, center, null$
    $)$ :
      $response \leftarrow HTTPResponse()$
      $citizen \leftarrow Citizen.queryset($
        $vaccination.pseudoUUID$
      $)$
      **try** $completeVaccination(citizen, response)$ :
        $print('Citizen successfully vaccinated')$
      **catch** $Exception\ e$ :
        $raise\ e$
    **catch** $Exception\ e$ :
      $raise\ e$

**Algorithm 10** Vaccination(citizen: map, center: map):

    $numberOfDoses \leftarrow max($
      $Vaccination.query(citizen).extract(doseNumber)$
    $)$
    **if** numberOfDoses == 2 **then**
        $raiseException('Citizen completely vaccinated')$
    **else**
      $this.doseNumber \leftarrow numberOfDoses + 1$
      $this.vaccinationID \leftarrow generateVaccinationID($
        $this.pseudoUUID,$
        $this.centerID,$
        $this.doseNumber$
      $)$
      $this.pseudoUUID \leftarrow citizen.pseudoUUID$
      $this.centerID \leftarrow center.centerID$
    **end if**

**Algorithm 11** generateVaccinationID(pseudoUUID: string, centerID: string, doseNumber):

    **return** $pseudoUUID + doseNumber + centerID$

The citizen is asked for the secret code linked to their profile and the PIN code of the area that they entered while registering on the portal. These two pieces of information are jointly used to establish the citizen's identity in the verification system. These are then queried and the pseudo-UUID associated with

**Algorithm 12** createVerificationPage(secretCode: string, PIN-Code: string, center: map, extraData: map):

    $citizen \leftarrow Citizen.query(secretCode, PINCode)$
    $r \leftarrow random((timestamp, citizen.pseudoUUID).$
        $encode(base = 10))$
    $challenge \leftarrow AES.encrypt(message = r,$
      $key = citizen.staticKey)$
    $this.solveChallenge($
      $challenge, r, citizen, center, extraData$
    $)$

**Algorithm 13** solveChallenge(challenge: string, r: string, citizen: map, center: map, extraData: map):

    $print(extraData)$
    $response \leftarrow HTTPResponse()$   //citizen response
    $challenge \leftarrow AES.decrypt(ciphertext = challenge,$
      $key = response.staticKey)$
    **if** attempt == r **then**
      **if** extraData == null **then**
        $vaccination \leftarrow newVaccination(citizen, center)$
        **return** $vaccination$
      **else**
        **return true**
      **end if**
    **else**
      $throwException('Verification failed')$
    **end if**

the citizen is extracted. A one-time-verification page is set up for the citizen for a small duration over the next few minutes. The verification challenge in this case is that the citizen needs to decrypt the 10-digit random number encrypted with their static key following a symmetric-key cryptography algorithm such as AES. The unique identifier (for example, the last 5 characters of the URL) is communicated with the citizen by the health official after which citizen goes to the page on their device and enters their static key and secret code. If the correct static key and secret code combination has been entered, the random number linked with the secret code for the verification gets matched with the decrypted random number and thus the verification is successful. The process is very much similar to the case of OTPs where passwords are sent to the registered phone numbers of the users, but require the overhead of storing the phone number in the system. In

**Algorithm 14** completeVaccination(citizen: map, response: map):

    $this.timestamp \leftarrow utc.now()$
    $this.vaccinator \leftarrow response.vaccinator$
    $this.healthData \leftarrow response.healthData$
    **return** $this.createVerificationPage($
      $citizen.secretCode, citizen.PINCode, null, this.data())$

our case, this overhead becomes a privacy concern as storing the phone number effectively reveals the citizen's identity to whoever accesses the phone number.

Once the verification is complete, a confirmation message is displayed to the citizen which they show the health officials there. The health officials can now see the previous vaccination history of the citizen by querying the secret code and PIN Code again to check if the citizen is eligible to get the next dose of the vaccine. Since the verification is now complete, the health officials can now proceed to administer the next vaccine dose to the citizen.

### D. Vaccination Confirmation and Certification

Once the vaccine dose has been administered to the citizen, the health official needs to mark the citizen as being vaccinated on the portal. Vaccination details such as name of the vaccine, the dose number, vaccinator's name, date and time of vaccination, etc. are entered into the system by the health official. For the purpose of further analysis, the citizen's health information is also recorded in the system such as post-vaccination reactions or any other symptoms that might be out of the ordinary. On entering all the details, the health officials signs it with the center's static key and a signing page is generated for the citizen on the portal. Similar to the step in the identity verification stage, the health official tells the last 5 characters of the page URL to the citizen. The citizen visits the page on their device and after checking the details entered by the health official, signs the same. Thus, both the citizen and health official have jointly confirmed the vaccination and a new vaccination entity is created in the database. A corresponding transaction is also generated and registered on the blockchain as described in the next section.

A vaccination certificate is generated for the citizen once the confirmation of vaccination has been done on the portal. The certificate is issued in the name of the citizen with details of the vaccination written in the same. It is signed by the government agency using its master key which establishes the authenticity of the certificate. The certificate can then be sent to the citizen as a soft copy, and can be used as a proof of vaccination by the citizen.

In the procedures and associated algorithms described above, it has been tried to maintain maximum privacy of the concerned citizen while ensuring all standard vaccination procedures are being followed. In some of the steps, a compromise has been reached between data privacy and the time consumed in the completion of the step such as the three-step identity verification process. As and when required, public key cryptographic techniques have been employed to preserve the confidentiality of citizens' data at steps which require interaction of both the citizens and associated third parties such as the health officials. The next section describes the post-vaccination stage of vaccination verification using the blockchain database employed in the system along with the necessary prerequisites to understand the same.

## IV. IMPLEMENTATION OF THE PROPOSED SYSTEM

The main use of a blockchain database in the proposed solution is to verify the transactions that have taken place in the procedures described above. The transactions refer to the state changes that have taken place in the system such as registration of a citizen and their vaccination. In order to store information about these transactions in the blockchain, it is necessary to define a format for the data that will be tracked through the transactions. We follow the guidelines specified by the IEEE Standard for Data Format for Blockchain Systems 2418.2TM-2020 [19] to design the data formats for the blocks and underlying transaction and entities in the blockchain.

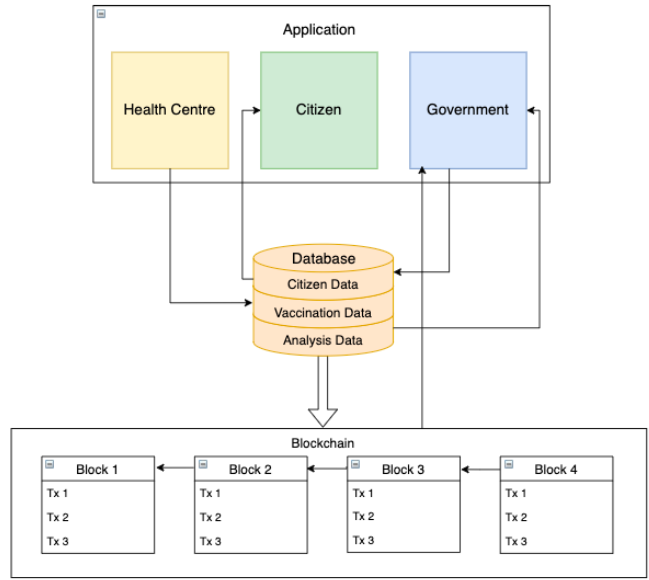

Fig. 7. High-Level System Overview

### A. Utilization of IEEE Standards in the Project [19]

The IEEE Standard for Data Format for Blockchain Systems 2418.2TM-2020 [19] specifies the need for outlining the data format for mainly 6 elements in a blockchain. These are the account data, block data, transaction data, entity data, contract data and configuration data. Out of these, the block, transaction, entity and contract data constitute the core elements of the blockchain. Multiple transaction objects may be encapsulated within a block, each of which captures the attributes of the entities involved in the block data as an Entity object and the business logic of the blockchain as a Contract object. Data corresponding to senders and receivers in block transactions are captured in the Account data entity. In this project, we have used and customized these data format specifications given in the standard to address the application's requirements. Implementation details specific to each data entity used in the project are given in the following subsection.

### B. Data Format

Based on the lines of the specifications as given in the standard, we have defined the data format for different data entities of the blocks suited to our application. We will primarily focus on describing the format of the account, transaction, entity and contract data with respect to the context in hand. Each of these different data entities are described under their respective sub-subsection below. The remaining data formats used in this application are given in the standard.

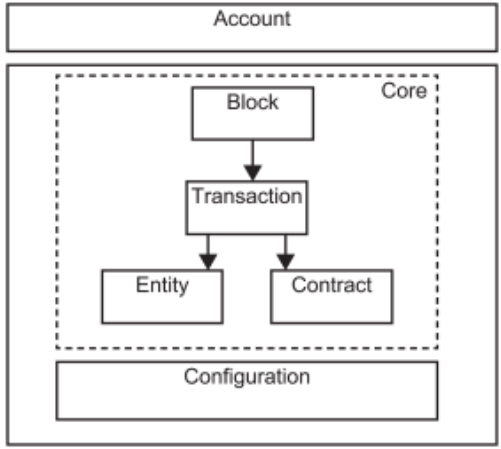

Fig. 8. Relationships between different data entities in the block data [19]

*1) Account Data:* The account data describes the details regarding the actual participating parties in the blockchain. They update the blockchain with new transaction which are associated with some account. In the proposed solution, the vaccination centers and government agencies are the accounts designated in the blockchain and each of them possess a public-private key pair for signing transactions and a list of account assets. In our application, the account assets for a vaccination center will contain the number of vaccine doses it possesses. This asset keeps on decreasing as new citizens get vaccinated at the center.

*2) Transaction Data:* The transaction data capture details about the business processes taking place in the application. A transaction may be a trading transaction or it may be a non-trading transaction as is the case with our system. The two main types of transactions that take place in the proposed solution are citizen registration and citizen vaccination. Each transaction is signed by one or more parties out of the vaccination center, citizen and government agency. The transaction timestamp tracks the exact time and date at which the transaction was executed which is used for temporal integrity verification of the transactions registered in the blockchain.

*3) Entity Data:* The static attributes of each transaction are described by the entity data associated with the transaction. This data generally captures details about the transaction sender and recipients, associated amounts and fees incurred. Additional details can also be added in this data for encompassing business details pertaining to the transaction. In the proposed solution, a citizen registration transaction is sent or initiated by a government agency and a vaccination transaction is initiated by a vaccination center. Recipient details are not included in these transactions. The number of newly registered and vaccinated citizens are captured through the amount data attributes in both transactions respectively and will be equal to 1 in all both the cases. Other details such as citizen pseudo-identity and vaccination details (as described in section II) are captured through the additional details and memo attributes of the entity data. One of the details stored in the memo is the primary key and hash value of the database entity corresponding to the transaction (eg. citizen profile, vaccination entity etc.), which will be used in data integrity verification later. The proposed solution doesn't charge fees for any new transactions and hence the fees attribute is dropped from the data format.

Fig. 9. Transaction Data Format of Citizen Registration and Vaccination transactions

Fig. 10. Entity Data Format of Citizen Registration and Vaccination transactions

*4) Contract Data:* The contract data summarizes the business logic of the procedures taking place in the system. The various processes taking place in the proposed solution as described in section III produce a number of transaction instances and the steps followed to generate these instances are captured in the contract data. It usually includes the contract code, version and storage information which collectively describe all the processes that produce the transaction instances. For example, the algorithms for registering a new citizen on the portal are coded in a programming language such as Solidity and this code is included under the code attribute of the contract data.

## C. Transaction Verification

The transactions that have been produced as a result of the different procedures taking place in the system, are organized as a Merkle tree in different blocks of the blockchain. Every block stores a hash value called Merkle Tree root in its block header. This value is the hash value of the root node of the Merkle tree formed by the transactions registered in the block. Although the block also maintains a list of the transactions in the form of an array, a Merkle tree is particularly helpful for the verification of the integrity of all the registered transaction. The leaf nodes of the tree are simply the hash values of the individual transactions. Every other node of the tree, is formed by hashing the concatenated hash values of its children. Thus, the root node is formed by the hash values of all the other nodes in the children and can be used to verify the integrity of all transactions in the block. In case a node has only one child, then this child is duplicated to make the node have two children and the rest of the process continues as above. In the Ethereum blockchain, the Merkle tree is replaced by a trie data structure called the Merkle-Patricia trie [20]. However, the idea behind the verification of data integrity of the underlying transactions remains similar.

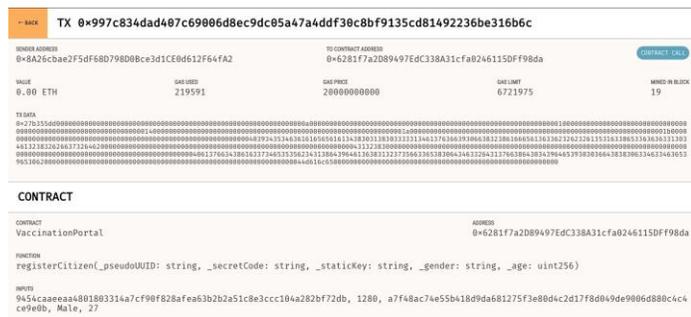

Fig. 11. Data for the transaction which registered a new citizen

In order to verify the transactions about the citizen registration and vaccinations, an authorized agency needs to connect to the blockchain database to read the transactions in the different blocks. In the proposed solution, this power is vested in the hands of the government agencies who will verify all the transactions taking place in the system. Each transaction has its corresponding database entity stored in a relational database managed by the portal. These are linked by the primary key of the database entity which is stored in the transaction and the transaction ID that is stored in the entity.

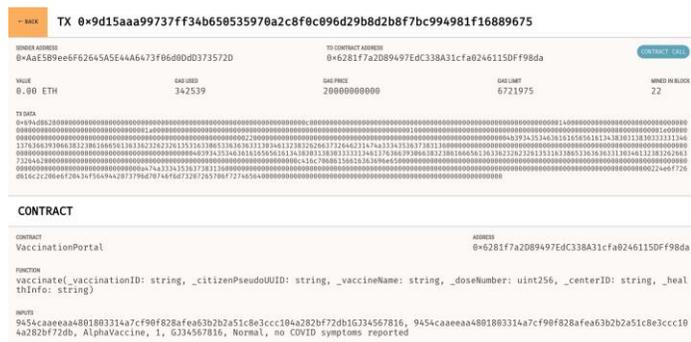

Fig. 12. Data for the transaction which registered a new vaccination for a citizen

A government agency might take up a particular vaccination instance from the database and check if it is consistent with the details it manifests such as existence of the citizen, number of vaccination doses given, etc. These can be found from other database entities as required to be checked by the question in hand. However, a database entity can be modified by system failure or by an accidental or deliberate attempt by a system administrator. In any case, the underlying transaction that was first registered when the database entity was first created, cannot be changed due to the immutability of the blockchain. Therefore, to check the integrity of the information found in the database, the government agency just needs to verify it with the corresponding transaction in the blockchain. This can be achieved by simply calculating the hash value of the database item and comparing it with the hash value stored in the memo attribute of the entity data in the corresponding transaction. This idea can also be extended to checking consistency in the number of vaccine doses administered by a center and the number of doses actually supplied to the center in the first place. Therefore, a large number of unethical practices such as vaccine black-marketing, hoarding, certificate loss and many others can be curbed by taking ideas from the proposed solution.

## V. CONCLUSION

The system proposed in this paper aims to implement a vaccination system which is not only efficient in the medical standards, but also preserves the privacy of the concerned citizens. Compromise of sensitive data such as UUIDs, address, date of birth, phone numbers has been kept to the utmost minimum which provides a transparent and identity-independent vaccination process. Some of the key features were inspired from the contemporary vaccination scenario in many countries of the world, while additional features such as pseudo-identity implementation, symmetric key encrypted verification pages, and the use of a blockchain to track vaccination history arose by the need of privacy and ethics in the system. The project has used the IEEE Standard for Data Format for Blockchain Systems to define the data entities used in the blockchain for the developed application. There still remain many aspects in the system which await research and innovation from the scientific community so that the system and its ideals can be further enhanced. From the perspective of a citizen, the system ensures that their personal data is not accidentally or deliberately exposed to some third-party and that they do not need to worry about the dependence of the vaccination statistics and analysis results on their individual identity. A government agency, on the other hand, has a transparent means of verifying the ongoing vaccination drives in its vicinity by virtue of the powerful properties of blockchain. Therefore, this system would be a very useful addition to the current vaccination scenario of the world and be of help to the civilian, medical and administrative communities. Some prominent future scopes of the proposed system are (i) to factor in demand-supply among vaccination centers; (ii) to integrate privacy-enabled customer satisfaction module; and (iii) compliance with privacy regulatory norms as per directives of the government and local authority.

ACKNOWLEDGEMENT

The authors would like to thank the IEEE Standard Education Grant programme for funding this project and providing the required motivation and guidance at all stages. Our sincere thanks to the IEEE Standards Association for the standards designed by them for easy use in research and development ventures. The student would like to express his gratitude and regards for his mentor Dr. Amit Dua for the continued guidance and support throughout the project.